\documentclass[preprint,showpacs,preprintnumbers,amsmath,amssymb,floatfix]{revtex4-1}

\usepackage{graphicx}
\usepackage{dcolumn}
\usepackage{bm}

\begin{document}


\title{Neutrino Capture Reactions on $^{40}$Ar} 

\author{Toshio Suzuki$^{1,2}$ and Michio Honma$^{3}$} 

\email{suzuki@chs.nihon-u.ac.jp}
\affiliation{$^{1}$ Department of Physics and Graduate School of Integrated 
Basic Sciences, College of Humanities and
Sciences, Nihon University     
Sakurajosui 3-25-40, Setagaya-ku, Tokyo 156-8550, Japan\\
%
$^{2}$ National Astronomical Observatory of Japan, 
Mitaka, Tokyo 181-8588, Japan\\
$^{3}$ Center for Mathematical Sciences, University of Aizu,
Aizu-Wakamatsu, Fukushima 965-8580, Japan}



\begin{abstract}
Gamow-Teller (GT) strength in $^{40}$Ar is studied by shell-model 
calculations with monopole-based universal intearction, which has tensor 
components of $\pi$+$\rho$-meson exchanges. Calculated GT strength is
found to be consistent with the experimental data obtained by recent
($p, n$) reactions. Neutrino capture cross sections on $^{40}$Ar for
solar neutrinos from $^{8}$B are found to be enhanced compared with
previous calculations.      
The reaction cross sections for multipoles other than $0^{+}$ and $1^{+}$
are obtained by random-phase approximation (RPA).  Their contributions
become important for neutrino energies larger than 50 MeV.  

\end{abstract}

\maketitle


\def\be{\begin{equation}}
\def\ee{\end{equation}}
\def\bea{\begin{eqnarray}}
\def\eea{\end{eqnarray}}
\def\br{\bf r}


    



\section{INTRODUCTION}

A liquid Argon detector has excellent potentialities to detect core-collapse
supernova neutrinos. A liquid Argon TPC (time projection chamber), proposed 
by the ICARUS Collaborations \cite{ICA}, can provide three-dimensional 
imaging of ionizing events. Recently, ArgoNeut Collaboration reported the 
first measurements of inclusive muon neutrino charged-current differential 
cross sections on argon \cite{Arg}.     
Here, we are interested in charged-current reactions on $^{40}$Ar induced
by solar and supernova neutrinos. Direct measurements of the charged-current
reaction cross sections on $^{40}$Ar are accessible by using a liquid Argon 
TPC detector and a spallation neutron source for neutrinos \cite{Cav}.

Recently, Gamow-Teller (GT) strength distribution for $^{40}$Ar $\rightarrow$ 
$^{40}$K was measured by using the ($p, n$) reaction up to an excitation
energy of $\sim$8 MeV \cite{Bhat}. The measured GT distribution is found to
be close to that obtained by $\beta$-decays of the mirror nucleus $^{40}$Ti
up to an excitation energy of $\sim$6 MeV \cite{Bhat98}. 
The GT strength obtained by a shell-model calculation \cite{Orm} is 
smaller than the observed strength    
As the neutrino capture reaction cross section on $^{40}$Ar is sensitive to
the magnitude and the distribution of the GT strength, it is a remarkable
progress if we can improve shell-model calculations to reproduce the 
experimental GT strength.  

As shell-model calculations of $^{40}$Ar involve both $sd$- and $pf$shells,
cross-shell correlations play important roles in 
the structure of the low-lying states in $^{40}$Ar.  
The choice of the cross-shell interaction is important and it will affect 
the GT strength distribution considerably.
Here, we adopt the monopole-based-universal interaction (VMU) \cite{VMU}
for the $sd$-$pf$ cross-shell part. The VMU has tensor components of
$\pi$+$\rho$-meson exchanges. A proper inclusion of the tensor force is
essential for a successful description of spin-dependent modes in nuclei.
The use of VMU in the $p$-$sd$ cross shell part of the interaction proved
to be successful in the description of Gamow-Teller transitions and 
magnetic moments in $p$-$sd$ shell nuclei \cite{Yuan}. 

In the next section, we discuss GT strength in $^{40}$Ar obtained by 
shell-model calculations and compare with experimental data. 
In Sect. III, cross sections of $^{40}$Ar ($\nu_e$, $e^{-}$) $^{40}$K 
reaction are obtained by a hybrid model; shell-model method is used for
the GT and isobaric analog (IA) transitions while random phase approximation
(RPA) is employed for other multipoles. Summary is given in Sect. IV.

\section{Gamow-Teller Strength in $^{40}$Ar with Monopole-based-
Universal Interaction}

\begin{figure*}[tbh]
\hspace{-8mm}
\includegraphics[scale=1.4]{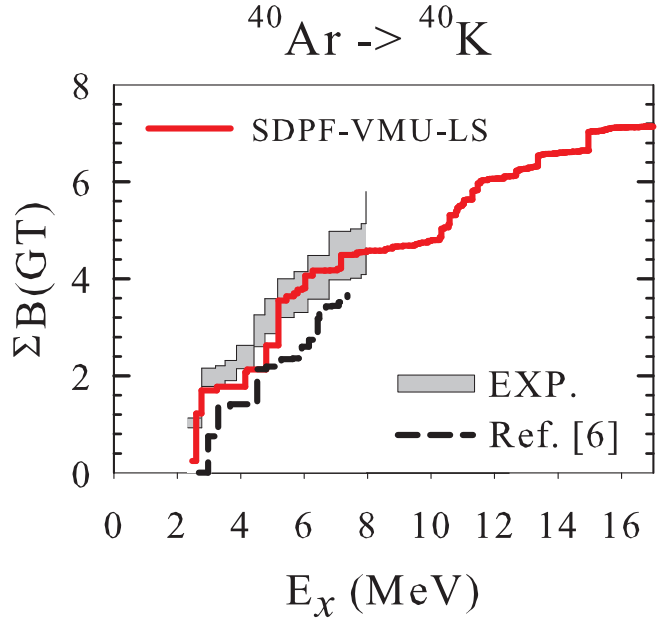}
\vspace{-5mm}
\caption{(Color on line) Cumulative sum of the GT strength for $^{40}$Ar 
$\rightarrow$ 
$^{40}$K up to excitation energies of $^{40}$K, $E_x$, obtained by 
the shell model calculation with the use of SDPF-VMU-LS.
The experimental data\cite{Bhat} are shown by shaded area. Calculated
values in Ref. \cite{Orm} are also shown by dashed line. 
\label{fig:fig1}}
\end{figure*}

We study charged-current neutrino-induced reaction on $^{40}$Ar, 
$^{40}$Ar ($\nu_e$, $e^{-}$) $^{40}$K, for neutrino energy below 100 MeV.
We adopt a hybrid model in which the Gamow-Teller (GT) transitions and
isobaric-analog (IA) transitions are evaluated by shell-model calculations
while contributions from other multipoles are treated by RPA.

We first discuss GT transitions in $^{40}$Ar. 
The sdpf-m \cite{Utsuno} and GXPF1J \cite{Honma} interactions are used 
for $sd$-shell and $pf$-shell, respectively, in the shell-model calculations.
The monopole-based-universal interaction (VMU) \cite{VMU}, which has the
tensor force of $\pi$+$\rho$-meson exchanges, is adopted for the $sd$-$pf$
cross-shell part. The two-body spin-orbit interaction due to meson exchanges
is also added to the cross-shell part of the interaction. 
This intercation will be referred as SDPF-VMU-LS. 
Configurations within 2$\hbar\omega$ excitations, $(sd)^{-2}(pf)^{2}$, 
are taken with a quenching factor of $f_{q}$ = $g_{A}^{eff}/g_{A}$ =0.775 
\cite{Orm} for the axial-vector coupling with $g_{A}$ =-1.263.    
Single-particle energies of $pf$-shell for $^{16}$O core are tuned to reproduce 
those of GXPF1J in $^{41}$Ca. No further tunings are made. 
The shell-model calculations are performed by using OXBASH \cite{OXB} and
MSHELL64 \cite{Mizu} program codes. 
Calculated ground state of $^{40}$K is 4$^{-}$ consistent with the experiment.
The energy of the first 1$^{+}$ state is also close to the experimental
excitation energy. In case of WBT interaction \cite{WB}, for example, it is shifted to
a higher energy region; excitation energy of the first 1$^{+}$ state is
calculated to be 4.14 MeV.      


Calculated cumulative sum of $B(GT)$, 
\begin{equation}
B(GT) = \frac{1}{2J_i +1} |<f || \sum_{k} f_q \sigma_k t_{k -} ||i>|^2,
\end{equation}

\noindent where $t_{-}|n>$ =$|p>$,
up to the excitation energy of $^{40}$K ($E_x$) are shown in Fig. 1 
as well as experimental values obtained by recent ($p$, $n$) reaction 
\cite{Bhat}.  
The experimental GT strength from the ($p, n$) reaction is consistent
with that obtained by $\beta$-decays from the mirror nucleus $^{40}$Ti 
\cite{Bhat98} except that the relative strengths for the lowest two strong 
transitions are reversed.  

The experimental cumulative sum of the GT strength is rather well 
described by SDPF-VMU-LS while the strength in Ref. \cite{Orm} is
smaller than the observed strength as shown in Fig. 1.
The GT strength of the first strong peak is larger than that of the second
strong peak in the present calculation, in agreement with the case of the
GT strength obtained by the ($p, n$) reaction.
The GT strength almost saturates around $E_x$ =18 MeV and the sum of the
GT strength amounts to be 7.21, which is equal to $f_q^{2}\times$12. 
The sum of the strength exhausts the Ikeda sum-rule value in the present 
configuration space.   

\section{Neutrino Capture Reaction Cross Sections on $^{40}$Ar}

\begin{figure*}[tbh]
\hspace{-5mm}
\includegraphics[scale=1.12]{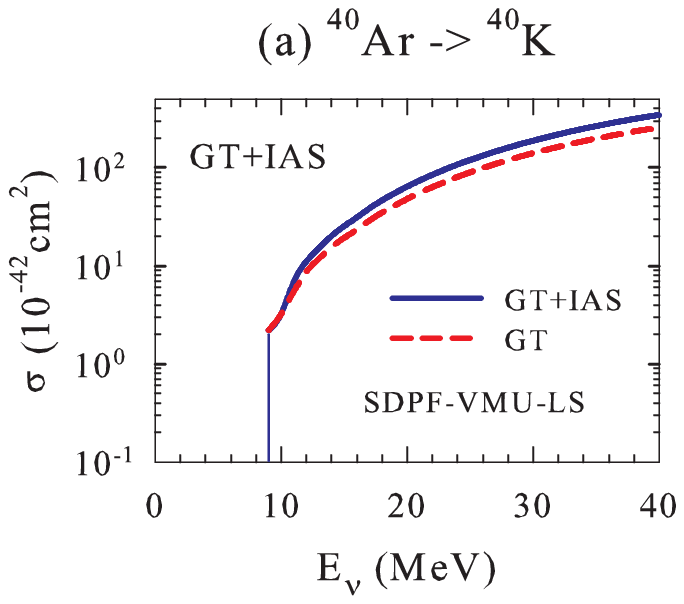}
\hspace*{-12mm}
\includegraphics[scale=1.12]{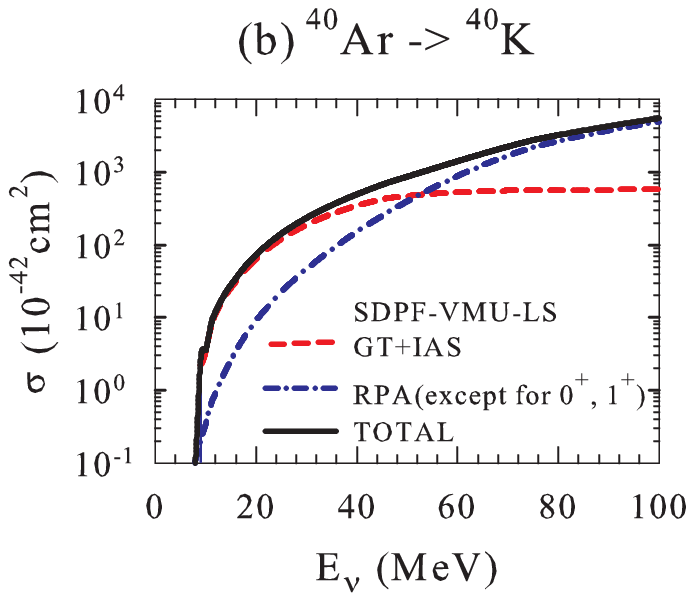}
\vspace{-12mm}
\caption{(Color on line) Calculated reaction cross sections for $^{40}$Ar
($\nu_e$, $e^{-}$) $^{40}$K. (a) Contributions from the GT and IA transitions
obtained by the shell-model calculations with the use of SDPF-VMU-LS are 
shown. (b) Contributions from multipoles except for $0^{+}$ and $1^{+}$ are
obtained by RPA calculations with the use of SGII \cite{SG}. Total cross
sections including all the multipoles are also shown by solid line.     
\label{fig:fig2}}
\end{figure*}

\begin{table*}[tbh]
\begin{center}
\caption{\label{tab:table1}
Calculated cross sections for $^{40}$Ar ($\nu_e$, $e^{-}$) $^{40}$K 
induced by solar $^{8}$B neutrinos for SDPF-VMU-LS and SDPF-VMU. 
Cross sections folded over the $^{8}$B neutrino spectrum \cite{Wint}
are given in unit of $10^{-43}$ cm$^{2}$.
Cross sections in Ref. \cite{Orm} are also given.}
\begin{tabular}{l|c|c|c}
\hline
Hamiltonian & GT & IA & GT+IA \\ 
\hline
SDPF-VMU-LS & 11.95 & 2.10 & 14.05 \\
SDPF-VMU & 10.99 & 2.10 & 13.10 \\
Ref. \cite{Orm} & 7.70 & 3.80 & 11.50 \\
\hline
\end{tabular}
\end{center}
\end{table*}

Neutrino-nucleus reaction cross sections are evaluated by using the 
multipole expansion of the weak hadronic currents,  
\begin{equation}
J_{\mu}^{C_{\mp}} = J_{\mu}^{V_{\mp}} + J_{\mu}^{A_{\mp}}
\end{equation}
for charge-exchange reactions ($\nu$, $\ell^{-}$) and ($\bar{\nu}, \ell^{+}$),
where $J_{\mu}^{V}$ and $J_{\mu}^{A}$ are vector and 
axial-vector currents, respectively.

The reaction cross sections induced by $\nu$ or $\bar{\nu}$ are given 
as follows \cite{Walecka}, 
\begin{eqnarray}
& &\left(\frac{d\sigma}{d\Omega}\right)_{\frac{\nu}{\bar{\nu}}} =
\frac{G^2\epsilon k}{4\pi^2}\frac{4\pi}{2J_i+1} 
\{\sum_{J=0}^{\infty} \{(1+\vec{\nu}\cdot\vec{\beta})\nonumber\\ 
& &\mid\langle J_f \parallel M_J \parallel J_i\rangle\mid^2 
+ [1-\hat{\nu}\cdot\vec{\beta}+2(\hat{\nu}\cdot\hat{q})(\hat{q}
\cdot\vec{\beta})]\nonumber\\
& &\mid\langle J_f\parallel L_J \parallel J_i\rangle
\mid^2
- \hat{q}\cdot(\hat{\nu}+\vec{\beta}) 2 Re \langle J_f \parallel
L_J \parallel J_i\rangle \nonumber\\
& &\langle J_f \parallel M_J \parallel J_i
\rangle^{\ast}\} 
+ \sum_{J=1}^{\infty} \{[1-(\hat{\nu}\cdot\hat{q})(\hat{q}\cdot
\vec{\beta})]\nonumber\\
& &(\mid\langle J_f \parallel T_{J}^{el} \parallel J_i
\rangle\mid^2 + \mid\langle J_f \parallel T_{J}^{mag} \parallel
J_i\rangle\mid^2 \nonumber\\
& &\pm \hat{q}\cdot(\hat{\nu}-\vec{\beta}) 2 Re [\langle J_f \parallel
T_{J}^{mag} \parallel J_i \rangle \nonumber\\ 
& &\langle J_f \parallel T_{J}^{el}
\parallel J_i \rangle^{\ast}])\} 
\end{eqnarray}
where $\vec{\nu}$ and $\vec{k}$ are neutrino and lepton momenta,
respectively, $\epsilon$ is the lepton energy, $\vec{q}=\vec{k}
-\vec{\nu}$, $\vec{\beta}=\vec{k}/\epsilon$, $\hat{\nu}=\vec{\nu}
/\mid\vec{\nu}\mid$ and $\hat{q}=\vec{q}/\mid\vec{q}\mid$. 
 
$G=G_{F}$cos$\theta_C$ with $G_{F}$ the Fermi coupling constant, 
$\theta_C$ is the Cabbibo angle, and the lepton is electron or positron 
for charge-exchange reactions. 
The Coulomb correction is taken into account by multiplying the Fermi 
function for charge-exchange reactions at low neutrino energires ($E_{\nu}$) 
while the correction by the effective momentum transfer approximation is 
used at higher energies.  
Lower value of the cross section obtained by the two methods 
is adopted at each $E_{\nu}$ \cite{Engel}.  
In Eq. (3), $M_{J}, L_{J}, T_{J}^{el}$ and $T_{J}^{mag}$ are
Coulomb, longitudinal, transverse electric and magnetic multipole
operators for vector and axial-vector currents

Calculated cross sections for $^{40}$Ar ($\nu_e$, $e^{-}$) $^{40}$K 
induced by GT and IA transitions are shown in Fig. 2(a).
Cross sections for $^{40}$Ar ($\nu_e$, $e^{-}$) $^{40}$K 
folded over $^{8}$B neutrino spectrum \cite{Wint} are given in Table I.
In Table I and Fig. 2, the cross sections are obtained with ICARUS 
condition, that is, electron energies are higher than 5 MeV.
The cross section for the GT transition is enhanced for SDPF-VMU-LS
compared to that of Ref. \cite{Orm} about by 55$\%$. When the two-body
spin-orbit interaction is switched off in the $p$-$sd$ cross-shell matrix
elements, which is referred as SDPF-VMU, the enhancement of the cross
section is about 43$\%$. The spin-orbit interaction is found to increase 
the cross section by about 9$\%$. 
Here, all the components of the GT transition, that is, the axial electric
dipole, the magnetic dipole, the axial Coulomb and the axial longitudinal
contributions are taken into acount while in Ref. \cite{Orm} only the
axial electric dipole term is considered. As for the IA transition, 
both the Coulomb ($M_{0^{+}}$) and the longitudinal ($L_{0^{+}}$) terms 
are taken into account. The inclusion of the longitudinal term reduces 
the total contributions in the IA transition as 
\begin{eqnarray}
& &|<1^{+}||M_{0^{+}}||0^{+}>|^2+|<1^{+}||L_{0^{+}}||0^{+}>|^2 \nonumber\\ 
&=&(\frac{q^2-\omega^2}{q^2})^2 |<1^{+}||M_{0^{+}}||0^{+}>|^2
\end{eqnarray}
This explains why the IA contribution in Ref. \cite{Orm}, where only the 
$M_{0^{+}}$ term is included, is smaller than the SDPF-VMU case. 
The cross section for GT+IA is also found to be enhanced for SDPF-VMU-LS
(SDPF-VMU) about by 22$\%$ (14$\%$) compared to ref. \cite{Orm}.

Cross sections for multipoles including up to $J$ =4 except for $0^{+}$ and 
$1^{+}$ are obtained
by RPA with the use of SGII interaction \cite{SG}. The same quenching
factor for $g_A$ as in the GT transitions is used. 
Calculated cross sections for GT, GT+IA, multipoles other than GT and IA, 
and the total cross section are shown in Fig. 2.   
The contributions from spin-dipole transitions and multipoles other than 
$0^{+}$ and $1^{+}$ are found to become important at $E_{\nu}\ge$ 50 MeV.

Calculated total cross section obtained here is found to be rather close to
that in Ref. \cite{Kolb} obtained by RPA calculations for all the
multipoles. The cross section by the present hybrid model is enhanced 
about by 20-40$\%$ compared to Ref. \cite{Kolb} at lower energies,
$E_{\nu}$ =20-40 MeV, where the GT contributions dominate. 
The calculated cross section in Ref. \cite{Chow} obtained by QRPA is
smaller than our cross sections as well as in Ref. \cite{Kolb} about by
three times. The GT distribution in Ref. \cite{Chow} is shifted toward 
higher energy region and quite a tiny strength is found below $E_{x}$
=8 MeV in contrast to the present case.

\section{Summary}

In summary, we have carried out shell-model calculations for GT and IA
transitions in $^{40}$Ar with the use of VMU interaction. The GT strength 
obtained is found to reproduce the experimental values rather well.   
The GT strength obtained here is larger than that in Ref. \cite{Orm}.  

Cross sections for the charged-current $\nu$-induced reaction, 
$^{40}$Ar ($\nu_e$, $e^{-}$) $^{40}$K, are evaluated by a hybrid model,
where contributions from the GT and IA transtions are obtained by
shell-model calculations while other multipoles are treated by RPA.
A better evaluation of the cross section for the GT contributions has
been achieved in the present work. 
An enhancement of the cross section for solar neutrinos from $^{8}$B 
is found compared to previous calculations. The contributions from
the GT and IA transitions dominate below $E_{x}\le$ 50 MeV while other
multipoles become important at higher energies. 
It is strongy recommended that
direct measurements of the cross section is carried out 
by using a liquid Ar TPC in near future.   
An accurate knowledge of the cross section is important for the advancement
of the studies of supernova neutrinos and neutrino oscillations. 

Here, we did not discuss neutral-current induced reactions on $^{40}$Ar
as the present shell-model configuration space is not large enough to
evaluate the magnetic dipole (M1) transition \cite{Li} and neutral-current 
induced transitions and the expansion of the configuration space beyond 
$(sd)^{-2}(pf)^2$ is difficult due to slow convergence of wave functions \cite{Orm}.
Study of neutral-current reactions with larger shell-model configurations
are left to future investigations.

This work has been supported in part by Grants-in-Aid for Scientific
Research (C) 22540290 
of the Ministry of Education, Culture, Sports,
Science and Technology of Japan. 


\end{document}